\documentclass[
    ,final            
  ]
  {aipproc}

\usepackage{bm}
\usepackage{amssymb}
\usepackage{amsmath}

\layoutstyle{6x9}

\begin{document}

\title{Mesoscopic Properties of \\ Molecular Folding and Aggregation Processes}


\author{Michael Bachmann}{
  address={Institut f\"ur Theoretische Physik and Centre for Theoretical Sciences (NTZ),
Universit\"at Leipzig, Postfach 100\,920, D-04009 Leipzig, Germany}
}

\author{Wolfhard Janke}{
  address={Institut f\"ur Theoretische Physik and Centre for Theoretical Sciences (NTZ),
Universit\"at Leipzig, Postfach 100\,920, D-04009 Leipzig, Germany}
}

\begin{abstract}
Protein folding, peptide aggregation and crystallization, as well as adsorption 
of molecules on soft or solid substrates have an essential 
feature in common: In all these processes, structure formation is guided by a 
collective, cooperative behavior of the molecular subunits lining up to build chainlike 
macromolecules. Proteins experience conformational transitions
related to thermodynamic phase transitions. For chains of finite length, an important 
difference of crossovers between conformational \mbox{(pseudo)phases} is, however, that 
these transitions are typically rather smooth processes, i.e., thermodynamic activity is not
necessarily signalized by strong entropic or energetic fluctuations. Nonetheless, in order to understand
generic properties of molecular structure-formation processes, the analysis of mesoscopic  
models from a statistical physics point of view enables 
first insights into the nature of conformational transitions in
small systems. Here, we review recent results obtained by means of sophisticated generalized-ensemble
computer simulations of minimalistic coarse-grained models.  
\end{abstract}

\maketitle


\section{Introduction}

At the atomic level, proteins have a complex chemical
structure which is formed and stabilized by the electronic properties of the 
atoms. Thus, the precise analysis of structure and dynamical behavior of molecules
require a detailed knowledge of the quantum mechanics involved. For molecular systems
of interest, where even the smallest molecules contain hundreds to thousands of atoms,
a quantum-mechanical analysis is usually simply impossible, a fortiori
if effects of the environment (e.g., an aqueous solution) are non-negligible. This
problem reflects the dilemma of ``realistic'' all-atom models which are based
on semiclassical assumptions and, consequently, depend on hundreds of parameters
mimicking quantum-mechanical effects. Actually, taken with sufficient care, the application
of such models is often inevitable if specific questions on atomic scales shall be 
investigated.

However, from a physics point of view, one may ask: Is it really necessary at all to employ all-atom 
models, if one is interested in \emph{generic} features of molecular mechanics which
typically anyway requires a cooperative action of larger subunits (monomers)? The
answer would be ``no'', if conformational transitions accompanying molecular 
structure-formation processes 
indeed exhibit similarities to thermodynamic phase transitions, in which case 
it should be possible to reveal general, qualitative properties by analyses of suitably simplified
models on mesoscopic scales~\cite{shak1,bj1}. 

After a few evolutionary remarks and introducing typical mesoscopic models, we 
eventually present results from studies of 
protein folding and peptide aggregation processes which
lead to the conclusion that characteristic features of the identified conformational
transitions are also relevant in corresponding natural structuring processes.

\section{The evolutionary aspect}

The number of different functional proteins encoded in the human DNA is of order 
100\,000~-- an extremely small number compared to the total number of 
possibilities: Recalling that 20 amino acids line up natural proteins and typical proteins
consist of $N\sim {\cal O}(10^2-10^3)$ amino acid residues, the number of
possible primary structures $20^N$ lies somewhere far, far above 
$\text{20}^\text{100}\sim \text{10}^\text{130}$. Assuming
all proteins were of size $N=100$ and a single folding event would take 1~ms, a sequential
enumeration process would need about $\text{10}^\text{119}$ years to generate structures of all sequences,
irrespective of the decision about their ``fitness'', i.e., the functionality and ability to
efficiently cooperate with other proteins in a biological system. Of course, one might argue 
that the evolution is a highly parallelized process which drastically increases the 
generation rate. So, we can ask the question, how many processes can maximally run in 
parallel. 

The universe contains of the order of $\text{10}^\text{80}$ protons. 
Assuming further that an average amino acid 
consists of at least 50 protons, a chain with $N=100$ amino acids has of the order ${\cal O}(10^3)$ 
protons, i.e.,
$\text{10}^\text{77}$ sequences could be generated in each millisecond (forgetting for the moment 
that some proton-containing 
machinery is necessary for the generation process and only a small
fraction of protons is assembled in earth-bound organic matter). The age of our universe is about
$\text{10}^\text{10}$ years (we also forget that the Earth is even about one order of magnitude younger)
or $\text{10}^\text{21}$~ms. 
Hence, about $\text{10}^\text{98}$ sequences could have been tested to date, if our 
drastic simplifications were right. But even this yet much too optimistic 
estimate is still noticeably smaller than the 
above mentioned reference number of $\text{10}^\text{130}$ possible sequences for a 100-mer. 

At 
least two conclusions can be
drawn from this crude analysis. One thing is that the evolutionary process of generating
and selecting sequences is ongoing as it is likely that only a small fraction of functional
proteins has been identified yet by nature. On the other hand, the existence of complex biological systems,
where hundreds of thousands different types of macromolecules interact efficiently, can only be 
explained by means of efficient evolutionary strategies of adaptation to environmental conditions
on Earth which dramatically changed through billions of years. Furthermore, the development
from primitive to complex biological systems leads to the conclusion that within the evolutionary 
process of protein design, particular patterns in the genetic code have survived over 
generations, while others were improved (or deselected) by recombinations, selections, 
and mutations. But the sequence question is only one side. Another regards the geometric 
structures of proteins which are directly connected to biological functionalities. The 
conformational similarity among human functional proteins is also quite surprising; only 
of the order of 1\,000 significantly different ``folds'' were identified~\cite{tang1}.

Since the conformation space is infinitely large because of the continuous degrees of 
freedom and the sequence space is also giant, the protein folding problem is typically
attacked from two sides: the \emph{direct folding problem}, where the amino acid sequence 
is given and the associated native, functional conformation has to be identified, and the 
\emph{inverse folding problem}, where one is interested in all sequences that fold into
a given target conformation. With these two approaches, it is, however, virtually impossible
to unravel evolutionary factors that led to the set of present functional proteins.
Only for very simple protein models, a comprising statistical analysis of
sequence and conformation spaces is possible.

\section{Simple approaches to coarse-grained modeling of proteins}

\begin{figure}
  \includegraphics[height=.3\textheight]{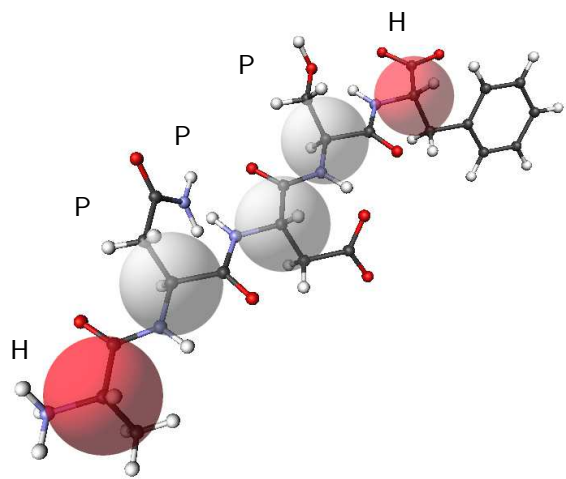}
  \caption{\label{fig:meso}Coarse-graining peptides in a ``united atom'' approach. Each amino acid is
contracted to a single ``C$^\alpha$'' interaction point. The effective distance
between adjacent, bonded interaction sites is about 3.8\,\AA. In the coarse-grained
hydrophobic-polar models considered here, the interaction sites have no steric
extension. The excluded volume is modeled via type-specific Lennard-Jones pair
potentials. In hydrophobic-polar (HP) peptide models, only hydrophobic (H)
and polar (P) amino acid residues are distinguished.}
\end{figure}
Coarse-graining of models, where 
relevant length scales are increased by reducing the number of microscopic degrees of 
freedom, has proven to be very successful in polymer science and protein folding~\cite{bj1}.
Although specificity is much more sensitive for proteins, since details (charges, polarity, etc.)
and differences of the amino acid side chains can have strong influences on 
the fold, also here mesoscopic approaches are of essential importance
for the basic understanding of conformational transitions affecting the folding process.
It is also the only possible approach for systematic analyses of basic problems
such as the evolutionarily significant question
why only a few sequences in nature are 
``designing'' and thus relevant for selective functions. On the other hand, what
is the reason why proteins prefer a comparative small set of target structures,
i.e., what explains the preference of designing sequences to fold into the 
\emph{same} three-dimensional structure? Many of these questions are 
still widely unanswered yet. Actually, the complexity of these questions requires a huge
number of comparative studies of complete classes of peptide sequences and structures
that cannot be achieved by means of computer simulations of microscopic models.
Currently only two approaches are promising. One is the bioinformatics approach of designing 
and scoring sequences and structures (and also possible combinations of receptors and 
ligands in aggregates), often based on data base scanning according to certain criteria.
Another, more physically motivated approach makes use of coarse-grained models, where
only a few specific properties of the monomers enter into the models. 
Since a characteristic feature of non-membrane proteins is to possess a compact hydrophobic core,
screened from the surrounding solvent by a shell of polar monomers, frequently only 
two types of amino acids are distinguished: hydrophobic (H) and polar (P) residues,
giving the class of corresponding models the name ``hydrophobic-polar'' (HP) models
(see Fig.~\ref{fig:meso}).

\subsection{The HP model for lattice proteins}

In  the simplest case, the HP peptide chain is a linear, self-avoiding chain 
of H and  P residues on a regular lattice~\cite{tang1,dill1}. Such models allow a comprising analysis
of both, the conformation \emph{and} sequence space, e.g., by exactly 
enumerating all combinatorial possibilities~\cite{schbj1}. Other important aspects in 
lattice model studies are the identification of lowest-energy conformations
of comparatively long sequences and the characterization of the folding 
thermodynamics~\cite{bj2}.

In the HP model,
a monomer of an HP sequence $\bm{\sigma}=(\sigma_1,\sigma_2,\ldots,\sigma_N)$ 
is characterized by its residual type
($\sigma_i=P$ for polar and $\sigma_i=H$ for hydrophobic residues), 
the position $1\le i\le N$ within the chain of length $N$, and the spatial position ${\bf x}_i$
to be measured in units of the lattice spacing. A conformation is then symbolized by the
vector of the coordinates of successive monomers, 
${\bf X}=({\bf x}_1,{\bf x}_2,\ldots,{\bf x}_N)$.
The distance between the $i$th and the $j$th monomer is denoted by $r_{ij}=|{\bf x}_i-{\bf x}_j|$.
The bond length between adjacent monomers in the chain is identical with the spacing of the
used regular lattice with coordination number $q$. These covalent bonds are thus
not stretchable.
A monomer and its nonbonded nearest neighbors may form so-called contacts.  
Therefore, the maximum number of contacts of a monomer within the chain is $(q-2)$ and
$(q-1)$ for the monomers at the ends of the chain. To account for the excluded volume,
lattice proteins are self-avoiding, i.e., two monomers cannot occupy the same 
lattice site. The total energy for an HP protein reads 
in energy units $\varepsilon_0$ (we set $\varepsilon_0=1$ in the following)
\begin{equation} 
\label{intro:eq:hpgen} 
E_\text{HP}=\varepsilon_0\sum\limits_{\langle i, j>i+1 \rangle} C_{ij}U_{\sigma_i\sigma_j},
\end{equation}
where $C_{ij}=(1-\delta_{i+1\,j})\Delta(x_{ij}-1)$ with 
\begin{equation}
\label{intro:eq:defDelta}
\Delta(z)=\left\{ \begin{array}{cl}
1, & \hspace{3mm} z=0,\\
0, & \hspace{3mm} z\neq 0
\end{array}\right. 
\end{equation}
is a symmetric $N\times N$ matrix called {\em contact map} and 
\begin{equation}
\label{intro:eq:intmatrix}
U_{\sigma_i\sigma_j}=\left(\begin{array}{cc}
u_{HH} & u_{HP}\\
u_{HP} & u_{PP} \end{array}\right)
\end{equation}
is the $2\times 2$ interaction matrix. Its elements $u_{\sigma_i\sigma_j}$ correspond to
the energy of $HH$, $HP$, and $PP$ contacts. For labeling purposes we shall adopt
the convention that $\sigma_i=0\,\hat{=}\, P$ and $\sigma_i=1\,\hat{=}\, H$.

In the simplest formulation~\cite{dill1}, only the attractive hydrophobic 
interaction is nonzero, $u^\text{HP}_{HH}=-1, u^\text{HP}_{HP}=u^\text{HP}_{PP}=0$.
Therefore, $U^\text{HP}_{\sigma_i\sigma_j}=-\delta_{\sigma_i H}\delta_{\sigma_j H}$. This 
parametrization has been extensively used to identify ground states of HP sequences, some of which are 
believed to show up qualitative properties comparable with realistic proteins whose 20-letter 
sequence was transcribed into the 2-letter code of the HP 
model~\cite{dill3,unger1,shak2,103lat,103toma}. 

This simple form of the standard HP model suffers, however, from the fact that the 
lowest-energy states are usually highly 
degenerate and therefore the number of designing sequences (i.e., sequences with unique 
ground state) 
is very small. Incorporating additional
inter-residue interactions~\cite{tang1}, symmetries are broken, degeneracies are smaller, and 
the number of designing sequences increases~\cite{schbj1}.

\subsection{A simple off-lattice generalization: The AB model}

Since lattice models suffer from undesired effects of the underlying lattice symmetries,
simple hydrophobic-polar \emph{off-lattice} models were defined.
One such model is the AB model, where, for historical reasons, 
A symbolizes hydrophobic and B polar
regions of the protein, whose conformations are modeled by polymer chains in continuum space
governed by effective bending energy and van der Waals interactions~\cite{still1}. These models
allow for the analysis of different
mutated sequences with respect to their folding characteristics. Here, the idea
is that the folding transition is a kind of 
pseudophase transition which can
in principle be described by one or a few order-like parameters. Depending on
the sequence, the folding process can be highly cooperative (downhill folding),
less cooperative depending on the height of a free-energy barrier 
(two-state folding), or even frustrating due to the existence of different 
barriers in a metastable regime (crystal or glassy phases)~\cite{baj1,ssbj1}. These characteristics
known from functional proteins can be recovered in the AB model, which is 
computationally much less demanding than all-atom formulations and thus 
enables throughout theoretical analyses.

We denote the spatial position of the $i$th monomer in a heteropolymer consisting 
of $N$ residues by ${\bf x}_i$, $i=1,\ldots,N$, and the vector connecting nonadjacent monomers $i$ and $j$ 
by ${\bf r}_{ij}={\bf x}_i-{\bf x}_j$. For covalent bond vectors, we set $|{\bf b}_i|\equiv |{\bf r}_{i\,i+1}|=1$. The bending angle 
between monomers $k$, $k+1$, and $k+2$ is $\vartheta_k$ 
($0\le \vartheta_k\le \pi$) and $\sigma_i=A,B$ symbolizes the type of the monomer. In the
AB model~\cite{still1}, the energy of a conformation is given by
\begin{equation}
\label{intro:eq:ab}
E_\text{AB} = \frac{1}{4}\sum\limits_{k=1}^{N-2}(1-\cos \vartheta_k)+
4\sum\limits_{i=1}^{N-2}\sum\limits_{j=i+2}^N\left(\frac{1}{r_{ij}^{12}}
-\frac{C(\sigma_i,\sigma_j)}{r_{ij}^6} \right),
\end{equation} 
where the first term is the bending energy and the sum runs over the $(N-2)$ bending angles of successive 
bond vectors. 
The second term partially competes with the bending barrier by a 
potential of Lennard-Jones type. It depends on the distance between monomers being nonadjacent 
along the chain and accounts for the influence of the AB sequence on the energy.
The long-range behavior is attractive for pairs of like monomers and repulsive for $AB$ pairs
of monomers:
\begin{equation}
\label{intro:eq:stillC}
C(\sigma_i,\sigma_j)=\left\{\begin{array}{cl}
+1, & \hspace{7mm} \sigma_i,\sigma_j=A,\\
+1/2, & \hspace{7mm} \sigma_i,\sigma_j=B,\\
-1/2,  & \hspace{7mm} \sigma_i\neq \sigma_j.\\
\end{array} \right.    
\end{equation}
The AB model is a
C$^\alpha$ type model in that each residue is represented by a single interaction site only, the
``C$^\alpha$ atom'' (see Fig.~\ref{fig:meso}). 
Thus, the natural dihedral torsional degrees of freedom of realistic
protein backbones are replaced by virtual bond and torsion angles. The large torsional barrier
of the peptide bond between neighboring amino acids is in the AB model effectively taken into
account by introducing the bending energy. 

Although this coarse-grained picture will obviously not be
sufficient to reproduce microscopic properties of specific realistic proteins, it qualitatively
exhibits, however, sequence-dependent features
known from nature, as, for example, tertiary folding
pathways characteristic for two-state folding, folding through intermediates, or 
metastability~\cite{ssbj1},
and two-state kinetics~\cite{kbj1}.

\section{Thermodynamics of heteropolymer folding}

For the analysis of conformational transitions accompanying the tertiary folding
behavior of lattice proteins, multicanonical chain-growth simulations~\cite{bj1,bj2} 
can be efficiently performed for the HP model. 
An example is the 42-mer with the sequence 
$PH_2PHP$\-$H_2$\-$P$\-$H$\-$P$\-$H$\-$P_2$\-$H_3$\-$P$\-$H$\-$P$\-$H_2$\-$P$\-$H$\-$P$\-$H_3$%
\-$P_2$\-$H$\-$P$\-$H$\-$P$\-$H_2$\-$P$\-$HPH_2P$ that forms a parallel
helix in the ground state. Originally, it was designed to serve as a lattice model of 
the parallel $\beta$ helix of {\em pectate lyase C}~\cite{yoder}. But there are additional
properties that make it an interesting and challenging
system. The ground-state energy
is known to be $E_\text{min}=-34$. In the simulations, 
the ground-state degeneracy was estimated
to be $g_0=3.9\pm 0.4$~\cite{bj1,bj2}, which is in
perfect agreement with the known value $g_0^\text{ex} = 4$ (except translational,
rotational, and reflection symmetries)~\cite{dill4}.
As we will see, there are two conformational transitions. At low temperatures,
fluctuations of energetic and structural quantities signalize a \mbox{(pseudo)}transition between the
lowest-energy states possessing compact hydrophobic cores and the regime of 
globular conformations, and at a higher temperature, there is another transition 
between globules  and random coils.     

\begin{figure}
\centerline {
\includegraphics[height=.3\textheight]{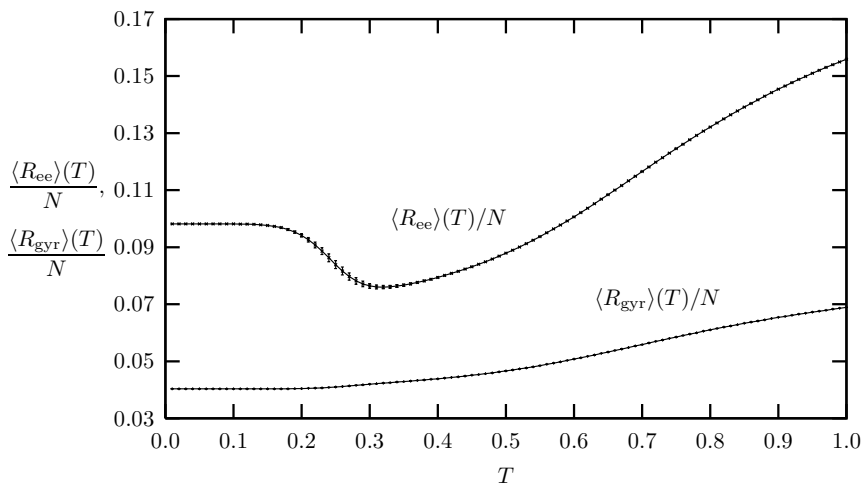}
}
\caption{\label{fig:hplat:struct42} 
Mean end-to-end distance $\langle R_\text{ee}\rangle $ and mean radius of gyration
$\langle R_\text{gyr}\rangle $ of the 42-mer.
}
\end{figure}

\begin{figure}
\centerline {
\includegraphics[height=.3\textheight]{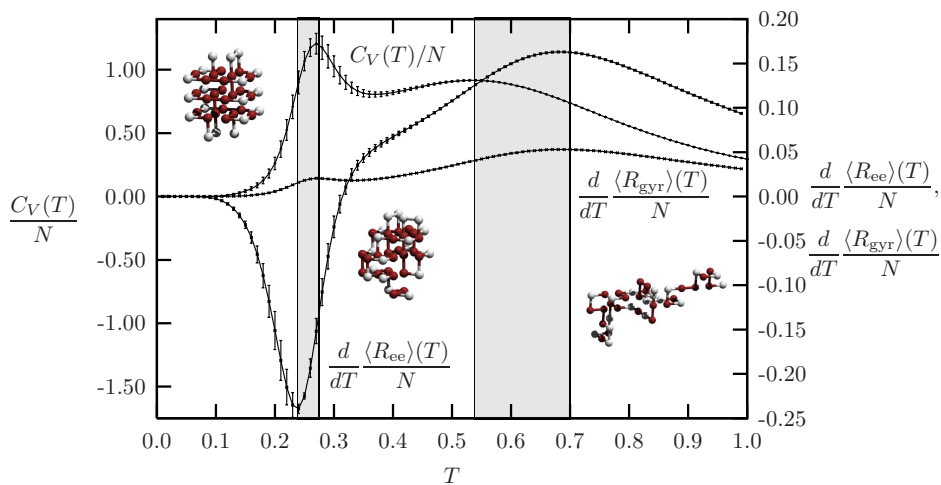}
}
\caption{\label{fig:hplat:42merfluct} 
Specific heat $C_V$ and derivatives w.r.t.\ temperature
of mean end-to-end distance $\langle R_\text{ee}\rangle$ and
radius of gyration $\langle R_\text{gyr}\rangle$ as functions of temperature for the 42-mer.
The ground-state~--~globule transition occurs between $T_0^{(1)}\approx 0.24$
and $T_0^{(2)}\approx 0.28$, while the globule~--~random coil transition takes
place between $T_1^{(1)}\approx 0.53$ and $T_1^{(2)}\approx 0.70$ (shaded areas).
}
\end{figure}

The average structural properties at finite temperatures can be characterized best 
by the mean end-to-end distance $\langle R_\text{ee} \rangle(T)$ and the mean radius 
of gyration $\langle R_\text{gyr} \rangle(T)$. 
Multicanonical chain-growth simulation results for $\langle R_\text{ee} \rangle(T)$ 
and $\langle R_\text{gyr} \rangle(T)$
of the 42-mer are shown in Fig.~\ref{fig:hplat:struct42}. 
The pronounced minimum in the end-to-end distance can be interpreted as an indication
of the transition between the lowest-energy states and globules: The
small number of ground states have similar and highly symmetric shapes 
(due to the reflection symmetry of the sequence) but the ends of the chain are 
polar and therefore they are not required to reside close to each other. 
Increasing the temperature allows the protein to fold into conformations different
from the ground states and contacts between the ends become more likely. Therefore, 
the mean end-to-end distance decreases and the protein has entered the globular
``phase''. Further increasing the temperature leads then to a disentanglement of
globular structures and random coil conformations with larger end-to-end distances dominate.      
In Fig.~\ref{fig:hplat:42merfluct}, we have plotted the specific heat $C_V(T)$ and the 
derivatives of the mean end-to-end distance and of the mean radius of gyration 
with respect to the temperature,
$d\langle R_\text{ee} \rangle/dT$ and $d\langle R_\text{gyr} \rangle/dT$.

Two temperature regions of conformational activity 
(shaded in gray),
where the curves of the fluctuating quantities exhibit extremal points, 
can clearly be separated. We estimate the temperature region of the
ground-state~--~globule transition to be within $T_0^{(1)}\approx 0.24$ and 
$T_0^{(2)}\approx 0.28$. The globule~--~random coil transition takes place
between $T_1^{(1)}\approx 0.53$ and $T_1^{(2)}\approx 0.70$. 

For high temperatures, random conformations are favored. In consequence, in the corresponding, rather 
entropy-dominated ensemble, the high-degenerate high-energy structures govern the thermodynamic 
behavior of the macrostates. A typical representative is shown as an inset in the 
high-temperature pseudophase in Fig.~\ref{fig:hplat:42merfluct}. Annealing the system (or, equivalently,
decreasing the solvent quality), the heteropolymer experiences a conformational transition
towards globular macrostates. A characteristic feature of these intermediary ``molten'' globules
is the compactness of the dominating
conformations as expressed by a small gyration radius. Nonetheless, the conformations
do not exhibit a noticeable internal long-range symmetry and behave rather like a fluid.
Local conformational changes are not hindered by strong free-energy barriers. The
situation changes by entering the low-temperature (or poor-solvent) conformational
phase. In this region, energy dominates over entropy and the effectively attractive 
hydrophobic force favors the formation of a maximally compact core of hydrophobic monomers. 
Polar residues are expelled to the surface of the globule and form a shell that
screens the core from the (fictitious) aqueous environment. 

The existence of the
hydrophobic-core collapse renders the folding behavior of a heteropolymer 
different from crystallization or amorphous transitions of homopolymers~\cite{vbj1}. The reason
is the disorder induced by the sequence of different monomer types. The hydrophobic-core
formation is the main cooperative conformational transition which accompanies
the tertiary folding process of a single-domain protein.    

\begin{figure}
\centerline {
\includegraphics[height=.6\textheight]{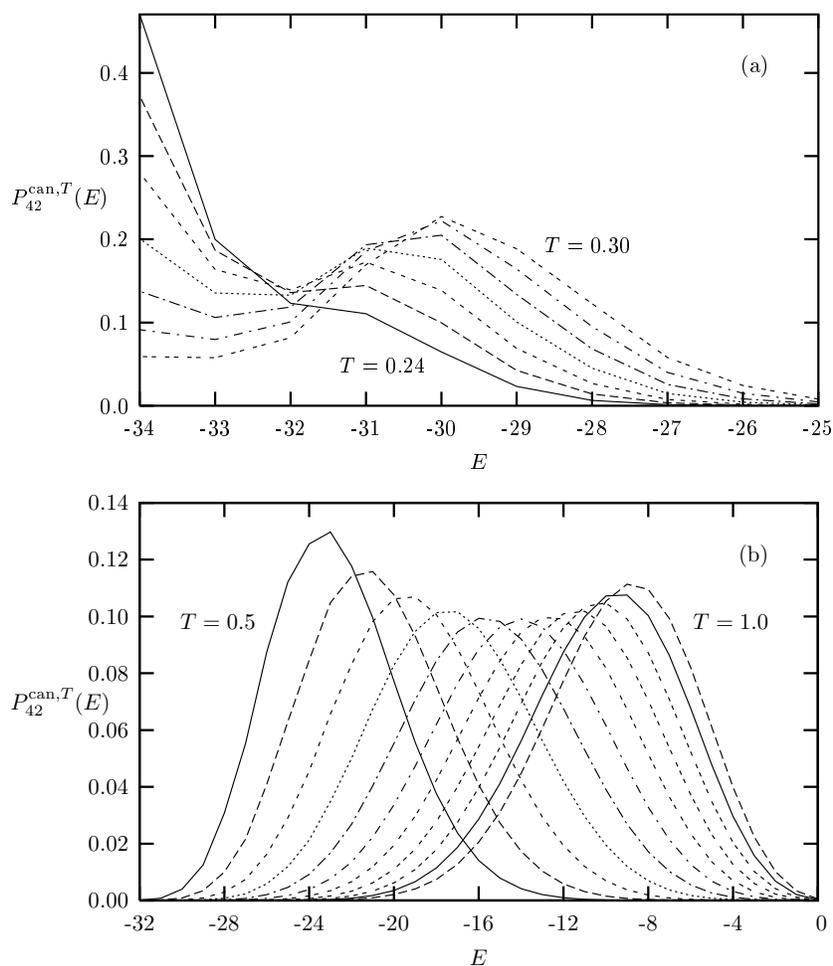}
}
\caption{\label{fig:hplat:42pE} 
Canonical distributions for the 42-mer at temperatures (a) $T=0.24,0.25,\ldots ,0.30$ close
to the ground-state -- globule transition region between $T_0^{(1)}\approx 0.24$
and $T_0^{(2)}\approx 0.28$, (b)
$T=0.50,0.55,\ldots ,1.0$. The high-temperature peak of the specific heat in Fig.~\ref{fig:hplat:42merfluct}
is near $T_1^{(1)}\approx 0.53$, but
at $T_1^{(2)}\approx 0.73$ the distribution has the largest width~\cite{bj2}.
Near this temperature, the
mean radius of gyration and the mean end-to-end distance 
(see Figs.~\ref{fig:hplat:struct42} and~\ref{fig:hplat:42merfluct}) have
their biggest slope.
}
\end{figure}

In Fig.~\ref{fig:hplat:42pE}
we have plotted the canonical distributions $p_{42}^{\text{can},T}(E)$ for different
temperatures in the vicinity of the two transitions. From \mbox{Fig.~\ref{fig:hplat:42pE}(a)}
we read off that the distributions possess two peaks
at temperatures within that region where the ground-state~--~globule 
transition takes place. This is interpreted 
as indication of a ``first-order-like'' transition, i.e., both types of macrostates
coexist in this temperature region~\cite{iba1}. The behavior in the
vicinity of the globule -- random 
coil transition is less spectacular as can be seen in \mbox{Fig.~\ref{fig:hplat:42pE}(b)}, 
and since the energy distribution 
shows up one peak only, this transition could be denoted as being ``second-order-like''.
The width of the distributions grows with increasing temperature until it has reached its 
maximum value which is located near $T\approx 0.7$. 
For higher temperatures, the distributions become narrower again~\cite{bj2}.   

\section{Protein folding is a finite-size effect}

Understanding protein folding by means of equilibrium statistical mechanics and thermodynamics is
a difficult task. A \emph{single} folding event of a protein cannot occur ``in equilibrium'' with its
environment. But protein folding is often considered as a folding/unfolding process
with folding and unfolding rates which are balanced in a stationary state that defines the
``chemical equilibrium''. Thus, the statistical properties of an infinitely long series of 
folding/unfolding cycles under 
constant external conditions (which are mediated by the surrounding solvent) can then also be
understood~--~at least in parts~-- 
from a thermodynamical point of view. In particular, folding and unfolding of a
protein are conformational transitions and one is tempted to simply take over 
the conceptual philosophy behind thermodynamic phase transitions, in particular known from
``freezing/melting'' and ``condensation/evaporation'' transitions of gases. But, such an
approach has to be taken with great care. Thermodynamic phase transitions occur only
in the thermodynamic limit, i.e., in infinitely large systems. A protein is, however, 
a heteropolymer uniquely
defined by its \emph{finite} amino acid sequence, which is actually comparatively short and cannot
be made longer without changing its specific properties. This is different for polymerized 
molecules (``homopolymers''),
where the infinite-length chain limit can be defined, in principle. The intensely studied
collapse or $\Theta$ transition between the random-coil and the globular phase is such a 
phase transition in the truest sense~\cite{deGennes1}, where a finite-size scaling towards the infinitely long
chain is feasible~\cite{grass1,vbj1}. In this case, also a classification of the phase transitions
into continuous transitions (where the latent heat vanishes and fluctuations exhibit power-law
behavior close to the critical point) and discontinuous transitions (with nonvanishing latent
heat) is possible.
 
For proteins (or heteropolymers with a ``disordered'' sequence), 
a finite-size scaling is impossible and so a classification of conformational transitions
in a strict sense. 
Nonetheless, cooperative conformational changes are often referred to as ``folding'', 
``hydrophobic-collapse'', ``hydrophobic-core formation'', or ``glassy'' transitions. All these
transitions are defined on the basis of certain parameters, also called ``order parameters''
or ``reaction coordinates'', but should not be confused with thermodynamic phase transitions.
The onset of finite-system transitions is also less spectacular: Their identification
on the basis of peaks and ``shoulders'' in fluctuations of energetic and structural quantities
and interpretation in terms of ``order parameters'' is a rather intricate procedure. Since
the different fluctuations do not ``collapse'' for finite systems, a unique transition
temperature can often not be defined. Despite a surprisingly high cooperativity, collective 
changes of protein conformations are not happening in a single step. 
As we have seen in Fig.~\ref{fig:hplat:42merfluct}, 
transition \emph{regions}
separate the ``pseudophases'', where random coils, maximally compact globules, or states
with compact hydrophobic core dominate~\cite{bj1,bj2,baj1}.

Although the lattice models are very useful in unraveling generic folding characteristics,
they suffer from lattice artifacts, which are, however, less relevant for long chains.
In order to obtain a more precise and thus finer resolved image of folding characteristics,
it is necessary to ``get rid of the lattice'' and to allow the coarse-grained protein to 
fold into the three-dimensional continuum.

\section{Tertiary protein folding channels from mesoscopic modeling}

Folding of linear chains of amino acids, i.e., bioproteins and synthetic peptides, is, 
for single-domain macromolecules, accompanied by the formation of secondary structures 
(helices, sheets, turns) and the tertiary hydrophobic-core collapse. While secondary
structures are typically localized and thus limited to segments of the peptide, the effective hydrophobic 
interaction between nonbonded, nonpolar amino acid side chains results in a global,
cooperative arrangement favoring folds with compact hydrophobic core and a surrounding polar
shell that screens the core from the polar solvent. 
Systematic analyses for unraveling general folding principles are extremely difficult 
in microscopic all-atom approaches, since the folding process is 
strongly dependent on the ``disordered'' sequence of amino acids 
and the native-fold formation is inevitably connected with, 
at least, significant parts of the sequence. Moreover,
for most proteins, the folding process is relatively slow (microseconds to seconds), which is  
due to a complex, rugged shape of the free-energy landscape~\cite{onuchic0,clementi2,onuchic1} with 
``hidden'' barriers, depending on sequence properties. 
Although there is no obvious system parameter that allows for a general 
description of the accompanying conformational transitions in folding processes 
(as, for example, the reaction coordinate in chemical reactions),
it is known that there are only a few classes of characteristic folding behaviors, mainly downhill
folding, two-state folding, folding through intermediates, and glass-like folding into metastable
conformations~\cite{du1,pande2,okamoto1,okamoto2,wolynes1,pande3,pitard1}.

Thus, if a classification of folding characteristics is useful at all, strongly simplified models
should reveal statistical~\cite{ssbj1} and kinetic~\cite{kbj1} 
pseudouniversal properties.
The reason why it appears useful to use a simplified, mesoscopic model like the AB model
is two-fold:
Firstly, it is believed that tertiary folding is mainly based on effective hydrophobic interactions
such that atomic details play a minor role. Secondly, systematic comparative folding studies for mutated 
or permuted sequences are computationally extremely demanding at the atomic level and are to date virtually
impossible for realistic proteins.
We will show in the following that by employing the  
AB heteropolymer model~(\ref{intro:eq:ab}) 
and monitoring a suitable simple angular similarity parameter 
it is indeed possible to identify 
different complex folding characteristics. 
The similarity parameter is defined as follows~\cite{baj1}:
\begin{equation}
\label{eq:ab:ov}
Q({\bf X},{\bf X'})=1 - d({\bf X},{\bf X'}). 
\end{equation}
With $N_b=N-2$ and $N_t=N-3$ being the respective numbers of 
bond angles $\Theta_i$ and torsional angles $\Phi_i$,
the angular deviation between the conformations is calculated according to
\begin{equation}
\label{eq:ab:dparam}
d({\bf X},{\bf X'})=\frac{1}{\pi(N_b+N_t)}\left[
\sum\limits_{i=1}^{N_b}d_b\left(\Theta_i,\Theta'_i\right)+
\min_{r=\pm}\left(\sum\limits_{i=1}^{N_t}d_t^{r}\left(\Phi_i,\Phi'_i\right)\right)\right],
\end{equation}
where
\begin{eqnarray}
\label{eq:dparamB}
d_b(\Theta_i,\Theta'_i)&=&|\Theta_i-\Theta'_i|,\nonumber\\
d_t^\pm(\Phi_i,\Phi'_i)&=&\text{min} \left(|\Phi_i\pm\Phi'_i|,2\pi-|\Phi_i\pm\Phi'_i| \right).
\end{eqnarray}
Here we have taken into account that the AB model is invariant under the reflection symmetry
$\Phi_i\to-\Phi_i$. Thus, it is not useful to distinguish between reflection-symmetric
conformations and therefore only the larger overlap is considered.
Since $-\pi\le \Phi_i\le \pi$ and $0\le\Theta_i\le\pi$, the overlap is unity, if all angles 
of the conformations ${\bf X}$ and ${\bf X'}$ coincide, else $0\le Q<1$. It should be noted that the average 
overlap of a random conformation with the corresponding reference state is for the sequences considered close to 
$\langle Q\rangle\approx 0.66$.
As a rule of thumb, it can be concluded that values $Q<0.8$ indicate weak or no significant similarity
of a given structure with the reference conformation.

\begin{figure}
\includegraphics[height=.8\textheight]{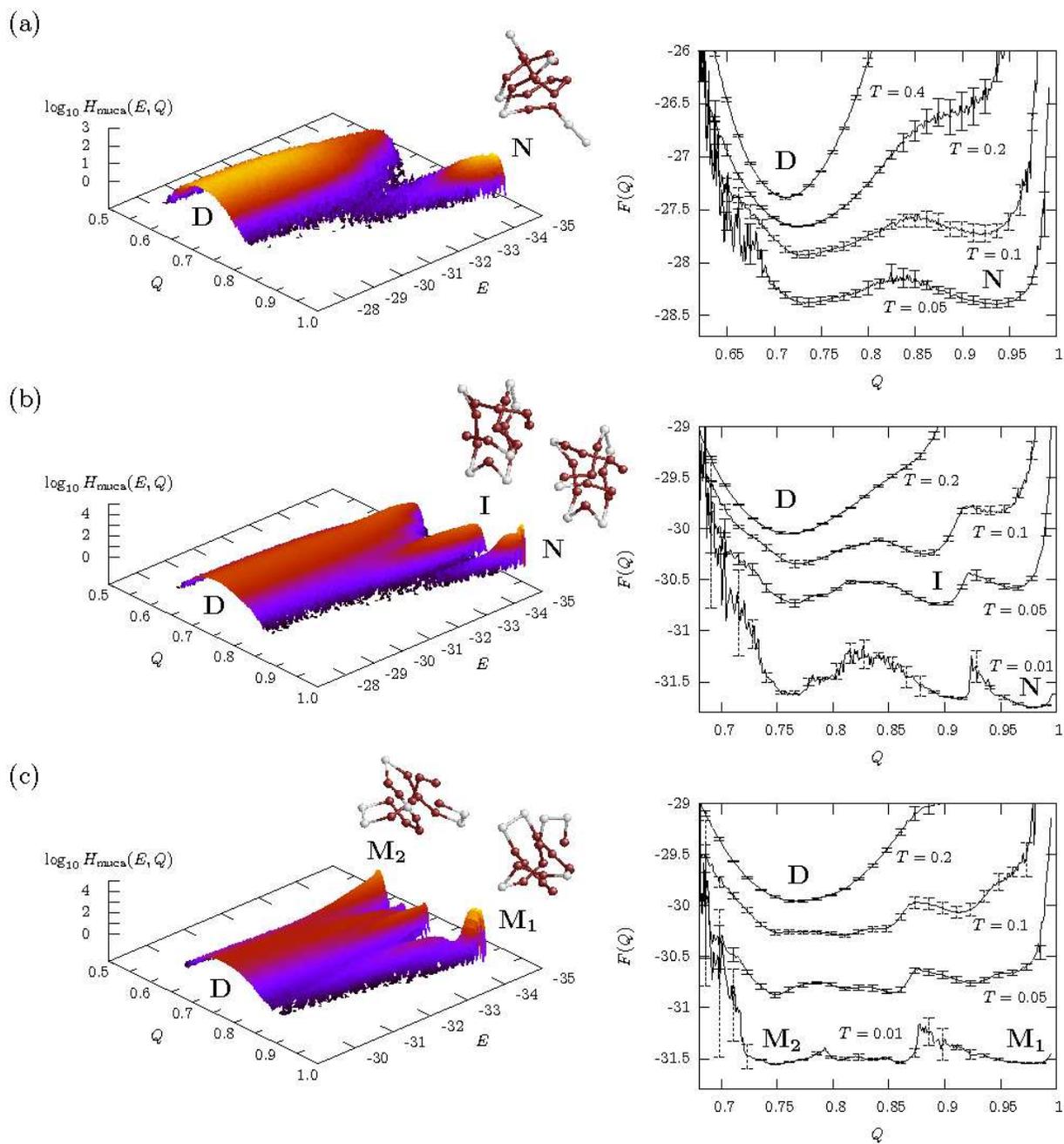}
\caption{\label{fig:class:fandh} Multicanonical histograms $H_\text{muca}(E,Q)$ 
of energy $E$ and angular overlap parameter $Q$ 
and free-energy landscapes $F(Q)$ at different temperatures for the three sequences 
(see Table~\ref{tab:ab:tabSEQ})
(a) S1, (b) S2, and (c) S3. 
The reference folds reside at $Q=1$ and $E=E_\text{min}$. Pseudophases are symbolized 
by D (denatured states), N (native
folds), I (intermediates), and M (metastable states). Representative conformations 
in intermediate and folded phases
are also shown~\cite{ssbj1}.}
\end{figure}

\begin{table}
\begin{tabular}{ll}
\hline
\tablehead{1}{l}{b}{Label} & \tablehead{1}{l}{b}{Sequence}   \\
\hline
S1 &  $BA_6BA_4BA_2BA_2B_2$ \\
S2 &  $A_4BA_2BABA_2B_2A_3BA_2$ \\
S3 &  $A_4B_2A_4BA_2BA_3B_2A$ \\
\hline
\end{tabular}
\caption{\label{tab:ab:tabSEQ}
Sequences of the heteropolymers compared with respect to their folding bahvior.
}
\end{table}
For the qualitative discussion of the folding behavior it is useful to consider the histogram
of energy $E$ and angular overlap parameter $Q$,
obtained from multicanonical simulations,
\begin{equation}
\label{eq:class:hmuca}
H_\text{muca}(E,Q) = \sum\limits_t\,\delta_{E,E({\bf X}_t)}\delta_{Q,Q({\bf X}_t,{\bf X}^{(0)})},
\end{equation} 
where the sum runs over all Monte Carlo sweeps $t$. 
In Figs.~\ref{fig:class:fandh}(a)--\ref{fig:class:fandh}(c), 
the multicanonical histograms $H_\text{muca}(E,Q)$ 
are plotted for the three sequences listed in Table~\ref{tab:ab:tabSEQ}.
Ideally, multicanonical sampling yields
a constant energy distribution 
\begin{equation}
\label{eq:class:hflat}
h_\text{muca}(E)= \int\limits_0^1dQ\,H_\text{muca}(E,Q) = \text{const.}
\end{equation}
In consequence, the distribution $H_\text{muca}(E,Q)$ can suitably be used to identify the folding channels, 
independently of temperature. This is more difficult with temperature-dependent canonical distributions 
$P^\text{can}(E,Q)$, which can, of course,
be obtained from $H_\text{muca}(E,Q)$ by a simple reweighting procedure, 
$P^\text{can}(E,Q)\sim H_\text{muca}(E,Q)g(E)\exp(-E/k_BT)$.
Nonetheless, it should be noted that, since there is a unique one-to-one 
correspondence between the average energy $\langle E\rangle$ and
temperature $T$, regions of changes in the monotonic behavior of $H_\text{muca}(E,Q)$ can 
also be assigned a temperature, where
a conformational transition occurs.
Interpreting the ridges of the probability distributions in the left-hand panel of 
Fig.~\ref{fig:class:fandh} as folding channels, it can 
clearly be seen that the heteropolymers exhibit noticeable differences in the folding behavior 
towards the native conformations (N). Considering natural proteins it would not be surprising that different
sequences of amino acids cause in many cases not only different native folds but also vary in their
folding behavior. Here we are considering, however, a highly minimalistic heteropolymer model and hitherto it
is not obvious that it is indeed possible to separate characteristic folding channels in this
simple model, but as Fig.~\ref{fig:class:fandh} demonstrates, in fact, it is. For sequence S1, we 
identify in Fig.~\ref{fig:class:fandh}(a) a typical 
two-state characteristics. Approaching from high energies (or high temperatures), 
the conformations in the ensemble of denatured conformations (D) 
have an angular overlap $Q\approx 0.7$, which means that there is 
no significant similarity with the reference
structure, i.e., the ensemble D consists mainly of unfolded peptides. For energies $E<-30$ a second branch opens. This channel (N) leads to the native conformation
(for which $Q=1$ and $E_\text{min}\approx -33.8$). The constant-energy distribution, where the main  
and native-fold channels D and N coexist, exhibits two peaks noticeably separated by a well. Therefore, the
conformational transition between the channels looks first-order-like, which is typical for
two-state folding. The main channel D contains the ensemble of unfolded conformations,
whereas the native-fold channel N represents the folded states. 

The two-state behavior is confirmed by analyzing the temperature dependence of the minima
in the free-energy landscape. The free energy as a function of the ``order'' parameter $Q$
at fixed temperature can be suitably defined as:
\begin{equation} 
\label{eq:class:freeE}
F(Q)=-k_BT\ln p(Q).
\end{equation}
In this expression,
\begin{equation}
\label{eq:class:pofq}
p(Q')= \int {\cal D}{\bf X}\, \delta(Q'-Q({\bf X},{\bf X}^{(0)}))\,e^{-E({\bf X})/k_BT}
\end{equation}
is related to the probability of finding a conformation with a given value of $Q$ in the canonical ensemble
at temperature $T$. The formal integration runs over all possible conformations ${\bf X}$. In the
right-hand panel of Fig.~\ref{fig:class:fandh}(a), 
the free-energy landscape at various temperatures is shown for sequence S1. At
comparatively high temperatures ($T=0.4$), only the unfolded states ($Q\approx 0.71$)
in the main folding channel D dominate. Decreasing the temperature, the second (native-fold) channel N begins 
to form ($Q\approx 0.9$), but the global free-energy minimum is still associated with the main channel. 
Near $T\approx 0.1$, both free-energy minima have approximately the same value, the folding transition
occurs. The discontinuous character of this conformational transition is manifest by the existence
of the free-energy barrier between the two macrostates. For even smaller temperatures, the native-fold-like
conformations ($Q>0.95$) dominate and fold smoothly towards the $Q=1$ reference conformation, 
which is the lowest-energy
conformation found in the simulation.

A significantly different folding behavior is noticed for the heteropolymer with sequence S2. The 
corresponding multicanonical histogram is shown in Fig.~\ref{fig:class:fandh}(b) and represents a folding
event through an intermediate macrostate. The main channel D bifurcates and a side channel I branches off
continuously. This branching is followed by the formation of a third channel N, which ends in the native fold. 
The characteristics of folding-through-intermediates
is also confirmed by the free-energy landscapes as shown for this sequence in 
Fig.~\ref{fig:class:fandh}(b) at different temperatures. Approaching from high temperatures, 
the ensemble of 
denatured conformations D ($Q\approx 0.76$) is dominant. Close to the transition temperature 
$T\approx 0.05$, the intermediary phase I is reached. The overlap of these intermediary conformations 
with the native fold is about $Q\approx 0.9$. Decreasing the temperature further below the native-folding
threshold close to $T=0.01$, the hydrophobic-core formation is finished and stable native-fold-like conformations with
$Q>0.97$ dominate (N).

The most extreme behavior of the three exemplified sequences is exhibited by the heteropolymer S3.
Figure~\ref{fig:class:fandh}(c) shows that 
the main channel D does not decay in favor of a native-fold channel. In fact, we observe both, the formation
of {\em two} separate native-fold channels M$_\text{1}$ and M$_\text{2}$. Channel
M$_\text{1}$ advances towards the $Q=1$ fold and M$_\text{2}$ ends up 
in a completely different conformation with approximately the same energy ($E\approx -33.5$). 
The spatial structures of these two conformations are noticeably different and their mutual overlap is
correspondingly very small, $Q\approx 0.75$. 
It should also be noted that the lowest-energy conformations in the main channel D
have only slightly larger energies than the two native folds. Thus, the folding of this heteropolymer
is accompanied by a very complex folding characteristics. In fact, this multiple-peak distribution
near minimum energies is a strong indication for metastability. A native fold in the natural sense
does not exist, the $Q=1$ conformation is only a reference state but the folding towards this
structure is not distinguished as it is in the folding characteristics of sequences S1 and S2.
The amorphous folding behavior is also seen in the free-energy
landscapes in Fig.~\ref{fig:class:fandh}(c). Above the folding transitions ($T=0.2$) the typical sequence-independent 
denatured conformations with $\langle Q\rangle \approx 0.77$ dominate (D). Then, in the annealing process, 
several channels are formed and coexist. The two most prominent channels (to which the lowest-energy
conformations belong that we found in the simulations) eventually lead for $T\approx 0.01$ to ensembles of macrostates
with $Q>0.97$ (M$_\text{1}$), and 
conformations with $Q<0.75$ (M$_\text{2}$). The lowest-energy conformation found in this regime 
is structurally different but energetically degenerate, if compared with the 
reference conformation.

\section{Statistical analyses of peptide aggregation}

\subsection{Pseudophase separation in polymeric nucleation processes}

Beside folding mechanisms, the aggregation of proteins
belongs to the biologically most relevant molecular structure formation processes. While the
specific docking between receptors and ligands is not necessarily accompanied by global
structural changes, protein folding and oligomerization of peptides are typically cooperative 
conformational transitions~\cite{gsponer1}. Proteins and their aggregates are comparatively small systems
and are often formed by only a few peptides. A very prominent example is the extracellular aggregation of the 
A$\beta$ peptide, which is associated with Alzheimer's disease. Following the amyloid hypothesis, 
it is believed that these aggregates 
 are neurotoxic, i.e., they are able to fuse into cell membranes of neurons and open calcium
ion channels. It is known that extracellular Ca$^{2+}$ ions intruding into a neuron can promote its
degeneration~\cite{lin1,quist1,lashuel1}.

Conformational transitions proteins experience
during structuring and aggregation are not phase transitions in the strict thermodynamic sense
and their statistical analysis is usually based on studies of signals exposed by energetic 
and structural fluctuations, as well as system-specific ``order'' parameters. In these studies, 
the temperature $T$ is considered as an adjustable,
external control parameter and, for the analysis of the pseudophase transitions, the peak structure 
of quantities such as the specific heat and the fluctuations of the gyration tensor components or 
``order'' parameter as functions of the temperature are investigated. The natural ensemble for 
this kind of analysis is the canonical ensemble, where the possible states of the 
system with energies $E$ are distributed according to the Boltzmann probability
$\exp(-E/k_BT)$, where $k_B$ is the Boltzmann constant. However, 
phase separation processes of small systems are accompanied 
by surface effects at the interface between the 
pseudophases~\cite{gross1,gross2}. This is reflected by the behavior
of the microcanonical entropy ${\cal S}(E)$, which exhibits a \textit{convex} monotony in the
transition region. Consequences are the backbending of the caloric temperature
$T(E)=(\partial {\cal S}/\partial E)^{-1}$, i.e., the \textit{decrease} of temperature 
with increasing system energy,
and the negativity of the microcanonical specific heat $C_V(E)=(\partial T(E)/\partial E)^{-1}=
-(\partial {\cal S}/\partial E)^2/(\partial^2 {\cal S}/\partial E^2)$. 
The physical reason is that the free energy 
balance in phase equilibrium requires the minimization of the interfacial surface and, therefore,
the loss of entropy~\cite{gross3,jbj1,jbj2}. 
A reduction of the entropy can, however, only be achieved by transferring energy 
into the system.  

It is a surprising fact that this so-called backbending effect is indeed observed in transitions
with phase separation. Although this phenomenon has already been known for a long time
from astrophysical systems~\cite{thirring1}, it has been widely ignored since then
as somehow ``exotic'' effect.
Recently, however, experimental evidence was found from melting studies of sodium clusters
by photofragmentation~\cite{schmidt1}. Bimodality and negative specific heats are also
known from nuclei fragmentation experiments and models~\cite{pichon1,lopez1}, as well as
from spin models on finite lattices which experience first-order transitions in the thermodynamic
limit~\cite{wj1,pleimling1}. This phenomenon is also observed in a large number of
other isolated finite model systems for evaporation and melting
effects~\cite{wales1,hilbert1}.

The following discussion of the aggregation behavior is based on 
multicanonical computer simulations of a mesoscopic hydrophobic-polar
heteropolymer model for aggregation based on the AB model~\cite{jbj1,jbj2}.

\subsection{Mesoscopic hydrophobic-polar aggregation model}

For studies of heteropolymer aggregation on mesoscopic scales, a novel
model is employed that is based on the hydrophobic-polar single-chain AB model~(\ref{intro:eq:ab}). 
As for modeling heteropolymer folding, we assume here that the tertiary folding
process of the individual chains is governed by hydrophobic-core formation in an aqueous 
environment. For systems of more than one chain, we further take into account 
that the interaction strengths
between nonbonded residues are independent of the individual properties of the chains the
residues belong to. Therefore, we use the same parameter sets as in the AB model for the 
pairwise interactions between residues of different chains. 
Our aggregation model reads~\cite{jbj1,jbj2}
\begin{equation}
\label{eq:agg:aggmod}
E=\sum\limits_{\mu} E_\text{AB}^{(\mu)}+\sum\limits_{\mu<\nu} 
\sum_{i_\mu,j_\nu}\Phi(r_{i_\mu j_\nu};\sigma_{i_\mu},\sigma_{j_\nu}),
\end{equation}
where $\mu,\nu$ label the $M$ polymers interacting with each other, and 
$i_\mu,j_\nu$ index the $N_{\mu,\nu}$ monomers of the respective $\mu$th and $\nu$th polymer.
The intrinsic single-chain energy of the $\mu$th polymer is given by [cf.\ Eq.~(\ref{intro:eq:ab})]
\begin{equation}
\label{eq:agg:abmod}
E_\text{AB}^{(\mu)}=\frac{1}{4}\sum\limits_{i_\mu}(1-\cos \vartheta_{i_\mu})+%
\!\!\sum\limits_{j_\mu>i_\mu+1}\Phi(r_{i_\mu j_\mu};\sigma_{i_\mu},\sigma_{j_\mu}),
\end{equation}
with $0\le \vartheta_{i_\mu}\le \pi$ denoting the bending angle between monomers 
$i_\mu$, $i_\mu+1$, and $i_\mu+2$.
The nonbonded inter-residue pair potential 
\begin{equation}
\label{eq:agg:phi}
\Phi(r_{i_\mu j_\nu};\sigma_{i_\mu},\sigma_{j_\nu})=
4\left[r_{i_\mu j_\nu}^{-12}-C(\sigma_{i_\mu},\sigma_{j_\nu})r_{i_\mu j_\nu}^{-6}\right]
\end{equation}
depends on the distance $r_{i_\mu j_\nu}$ between the residues, and on their type,
$\sigma_{i_\mu}=A,B$. The long-range behavior is attractive for 
like pairs of residues [$C(A,A)=1$, $C(B,B)=0.5$] and repulsive otherwise [$C(A,B)=C(B,A)=-0.5$]. 
The lengths of all virtual peptide bonds are set to unity. 

Employing this model, we study in the following thermodynamic properties of the aggregation 
of oligomers with the Fibonacci sequence 
F1:~$AB_2AB_2ABAB_2AB$ over the whole energy and temperature 
regime. 

\subsection{Order parameter of aggregation and fluctuations}

In order to distinguish between the fragmented and the aggregated regime, we introduce 
the ``order'' parameter~\cite{jbj1,jbj2}
\begin{equation}
\Gamma^2=\frac{1}{2M^2}\sum_{\mu,\nu=1}^{M} {\bf d}_\text{per}^2\left({\bf r}_{\text{COM},\mu},
{\bf r}_{\text{COM},\nu}\right),
\end{equation}
where the summations are taken over the minimum distances 
${\bf d}_\text{per}=\left(d_\text{per}^{(1)},d_\text{per}^{(2)},d_\text{per}^{(3)}\right)$
of the respective  
centers of mass of the chains (or their periodic continuations).  
The center of mass of the $\mu$th chain in a box with periodic boundary conditions
is defined as ${\bf r}_{\text{COM},\mu}=\sum_{i_\mu=1}^{N_\mu} \left[
{\bf d}_\text{per}\left({\bf r}_{i_\mu},{\bf r}_{1_\mu}\right)+{\bf r}_{1_\mu}\right]/N_\mu$, where
${\bf r}_{1_\mu}$ is the coordinate vector of the first monomer and serves as a reference coordinate 
in a local coordinate system. 

The aggregation parameter is to be considered 
as a qualitative measure; roughly, fragmentation corresponds to large values of $\Gamma$, aggregation
requires the centers of masses to be close to each other, in which case $\Gamma$ is comparatively small.
Despite its qualitative nature, it turns out to be a surprisingly manifest indicator for the 
aggregation transition and allows even a clear discrimination of different aggregation pathways, as
will be seen later on.

According to the Boltzmann distribution, we define canonical expectation values 
of any observable $O$ by
\begin{equation}
\langle O\rangle(T)=\frac{1}{Z_\text{can}(T)}
\prod\limits_{\mu=1}^M\left[\int {\cal D}{\bf X}_\mu\right]O(\{{\bf X}_\mu\})
e^{-E(\{{\bf X}_\mu\})/k_BT},
\end{equation}
where the canonical partition function $Z_\text{can}$ is given by
\begin{equation}
Z_\text{can}(T)=\prod\limits_{\mu=1}^M\left[\int {\cal D}{\bf X}_\mu\right]e^{-E(\{{\bf X}_\mu\})/k_BT}.
\end{equation}
Formally, the integrations are performed over all possible conformations ${\bf X}_\mu$ of the $M$
chains.

Similarly to the specific heat per monomer
$c_V(T)=d\langle E\rangle/N_\text{tot}dT=(\langle E^2\rangle-\langle E\rangle^2)/N_\text{tot}k_BT^2$
(with $N_\text{tot}=\sum_{\mu=1}^M N_\mu$) 
which expresses the thermal fluctuations of energy, the temperature derivative 
of $\langle\Gamma\rangle$ per monomer,
$d\langle\Gamma\rangle/N_\text{tot}dT=(\langle \Gamma E\rangle-\langle\Gamma\rangle\langle E\rangle)/N_\text{tot}k_BT^2$, is 
a useful indicator for cooperative behavior of the multiple-chain system. 
Since the system size
is small~-- the number of monomers $N_\text{tot}$ as well as the number of chains $M$~-- 
aggregation transitions,
if any, are expected to be signalized by the peak structure of the fluctuating quantities as functions
of the temperature. This requires the temperature to be a unique external control parameter which
is a natural choice in the canonical statistical ensemble. Furthermore, this is a typically easily
adjustable and, therefore, convenient parameter in experiments.  
However, aggregation is a phase separation process and, since the system is small, there is no uniform
mapping between temperature and energy~\cite{jbj1,jbj2}. 
For this reason, the total system energy is the more appropriate
external parameter. Thus, the microcanonical interpretation will turn out to be the more favorable 
description, at least in the transition 
region.

\subsection{Canonical and microcanonical interpretation}

\begin{figure}
\includegraphics[height=.3\textheight]{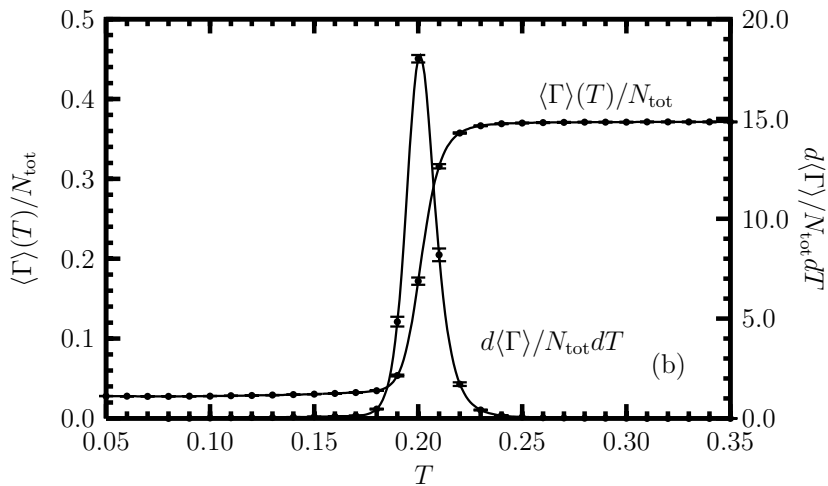}
\caption{\label{fig:agg:canon2x13.1} 
Aggregation parameter $\langle \Gamma\rangle/N_\text{tot}$ and 
fluctuations $d\langle\Gamma\rangle/N_\text{tot}dT$ as functions
of the temperature.
}
\end{figure}
In our aggregation study of the 2$\times$F1 system we obtain from the canonical analysis
a surprisingly clear picture of the aggregation transition.
In Fig.~\ref{fig:agg:canon2x13.1}, the 
temperature dependence of the mean aggregation order parameter $\langle \Gamma\rangle$ and 
the fluctuations of $\Gamma$ are shown. The aggregation transition is signalized by a very sharp
peak and we read off an aggregation temperature close to $T_\text{agg}\approx 0.20$. The 
aggregation of the two peptides is a single-step process, in which the formation of the 
aggregate with a common compact hydrophobic core governs the folding behavior of the individual
chains. Under such conditions, folding and binding are not separate processes. This is
different if the intrinsic polymeric forces are stronger than the binding affinity. In this 
case the already folded molecule simply docks at the active site of a target without changing its global
conformation. These two principal scenarios are also observed in molecular adsorption
processes at solid substrates. This provides another example where mesoscopic
models prove extremely useful in order to reveal the structural phases of adsorbed and desorbed 
conformations in dependence of external parameters such as
solvent quality and temperature~\cite{adbj1,adbj2}.

\begin{figure}
\includegraphics[height=.3\textheight]{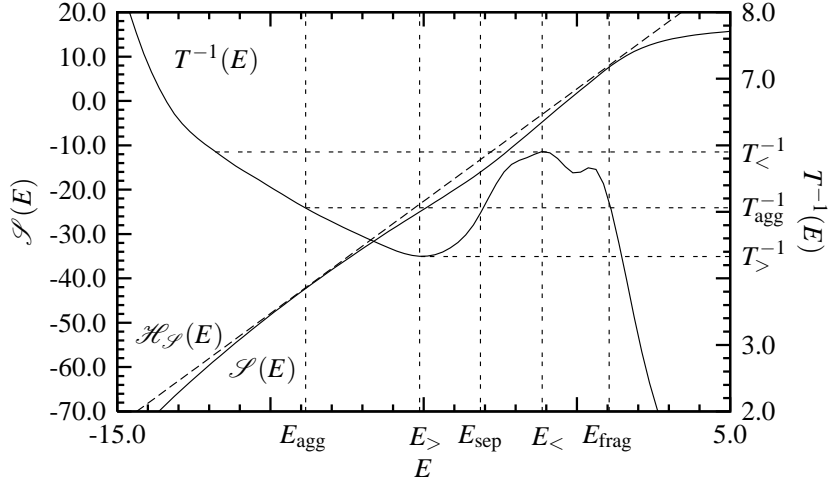}
\caption{\label{fig:agg:entropy} 
Microcanonical Hertz entropy ${\cal S}(E)$ of the 2$\times$F1 system,
concave Gibbs hull ${\cal H}_{\cal S}(E)$, and inverse caloric temperature $T^{-1}(E)$ as functions of 
energy. The phase separation regime ranges from $E_\text{agg}$ to $E_\text{frag}$. 
Between $T^{-1}_<$ and $T^{-1}_>$, the temperature is no suitable external control parameter and
the canonical interpretation is not useful: The inverse caloric temperature $T^{-1}(E)$ exhibits an
obvious backbending in the transition region. Note the second, less-pronounced backbending
in the energy range $E_<<E<E_\text{frag}$.
}
\end{figure}

In the microcanonical analysis of peptide-peptide aggregation, 
the system energy $E$ is kept (almost) fixed and treated as
an external control parameter. The system can only take macrostates with energies in the interval
$(E,E+\Delta E)$ with $\Delta E$ being sufficiently small to satisfy $\Delta G(E)=g(E)\Delta E$,
where $\Delta G(E)$ is the phase-space volume of this energetic shell. In the limit $\Delta E\to 0$,
the total phase-space volume up to the energy $E$ can thus be expressed as 
\begin{equation}
\label{eq:agg:psvol}
G(E)=\int_{E_\text{min}}^EdE'\,g(E'). 
\end{equation}
Since $g(E)$ is positive for all $E$, $G(E)$ is a monotonically 
increasing function and this quantity is suitably related to the microcanonical 
entropy ${\cal S}(E)$ of the system. In the definition of Hertz, 
\begin{equation}
\label{eq:agg:hertz}
{\cal S}(E)=k_B\ln\,G(E).
\end{equation}
Alternatively, the entropy is often directly related to the density of states $g(E)$ and defined
as 
\begin{equation}
\label{eq:agg:ent}
S(E)=k_B\ln\,g(E). 
\end{equation}
The density of states exhibits a decrease much faster than exponential towards the low-energy 
states. For this reason, the phase-space volume at
energy $E$ is strongly dominated by the number of states in the energy shell $\Delta E$. Thus 
$G(E)\approx \Delta G(E)\sim g(E)$ is directly related to the density of states. This virtual
identity breaks down in the higher-energy region, where $\ln\, g(E)$ is getting flat~-- in our case far
above the energetic regions being relevant for the discussion of the aggregation transition
(i.e., for energies $E\gg E_\text{frag}$, see Fig.~\ref{fig:agg:entropy}).
Actually, both definitions of the entropy lead to virtually
identical results in the analysis of the aggregation transition~\cite{jbj1,jbj2}.
The (reciprocal) slope of the microcanonical entropy fixes the temperature scale and the corresponding 
caloric temperature is then defined via 
$T(E)=(\partial {\cal S}(E)/\partial E)^{-1}$ for fixed volume $V$ and particle number $N_\text{tot}$.

As long as the mapping between the caloric temperature $T$ and the system energy $E$ is bijective,
the canonical analysis of crossover and phase transitions is suitable since the temperature can be
treated as external control parameter. For systems, where this condition is not satisfied, however, 
in a standard canonical analysis one may easily miss a physical effect accompanying 
condensation processes: Due to surface effects (the formation of the contact surface between the peptides
requires a rearrangement of monomers in the surfaces of the individual peptides), 
additional energy does not necessarily lead to
an increase of temperature of the condensate. Actually, the aggregate can even become colder. 
The supply of additional energy supports the fragmentation of parts of the aggregate, but this is 
overcompensated by cooperative processes of the particles aiming to reduce the surface tension.
Condensation processes are phase-separation processes and as such aggregated and fragmented phases
coexist. Since in this phase-separation region $T$ and $E$ are not bijective, this phenomenon is
called the ``backbending effect''. The probably most important class of systems exhibiting this 
effect is characterized by their smallness and the capability to form aggregates, depending on the 
interaction range. The fact that this effect could be indirectly observed in
sodium clustering experiments~\cite{schmidt1} gives rise to the hope
that backbending could also be observed in aggregation processes of small peptides.

Since the 2$\times$F1 system apparently belongs to this class, the backbending effect is
also observed in the aggregation/fragmentation transition of this system. This is shown 
in Fig.~\ref{fig:agg:entropy}, where the microcanonical entropy ${\cal S}(E)$ is plotted as function 
of the system energy. The phase-separation region of aggregated and fragmented conformations
lies between $E_\text{agg}\approx -8.85$ and $E_\text{frag}\approx 1.05$.
Constructing the concave Gibbs
hull ${\cal H}_{\cal S}(E)$ by linearly connecting ${\cal S}(E_\text{agg})$ and ${\cal S}(E_\text{frag})$ 
(straight dashed line in Fig.~\ref{fig:agg:entropy}), the entropic deviation due to surface effects is
simply $\Delta {\cal S}(E)={\cal H}_{\cal S}(E)-{\cal S}(E)$. The deviation is maximal for  
$E=E_\text{sep}$ and $\Delta {\cal S}(E_\text{sep})\equiv \Delta {\cal S}_\text{surf}$ is the
surface entropy. The Gibbs hull also defines the aggregation transition temperature
\begin{equation}
\label{eq:agg:tagg}
T_\text{agg}=\left(\frac{\partial {\cal H}_{\cal S}(E)}{\partial E}\right)^{-1}.
\end{equation}
For the 2$\times$F1 system, we find $T_\text{agg}\approx 0.198$, which is virtually
identical with the peak temperature of the aggregation parameter fluctuation 
(see Fig.~\ref{fig:agg:canon2x13.1}). 

The inverse caloric temperature $T^{-1}(E)$ is also plotted into Fig.~\ref{fig:agg:entropy}. For a fixed
temperature in the interval $T_< <T< T_>$ ($T_< \approx 0.169$ and $T_> \approx 0.231$), 
different energetic macrostates coexist. This is a consequence
of the backbending effect. Within the backbending region, the temperature decreases with increasing
system energy. The horizontal line at $T^{-1}_\text{agg}\approx 5.04$ is the Maxwell construction, i.e.,
the slope of the Gibbs hull ${\cal H}_{\cal S}(E)$. Although the transition seems to have similarities
with the van der Waals description of the condensation/evaporation transition of gases~-- the ``overheating''
of the aggregate between $T_\text{agg}$ and $T_>$ (within the energy interval 
$E_\text{agg}<E<E_>\approx -5.13$) 
is as apparent as the ``undercooling'' of the fragments between $T_<$ and $T_\text{agg}$ 
(in the energy interval $E_\text{frag}>E>E_<\approx -1.13$)~-- it is important 
to notice that in contrast to
the van der Waals picture the backbending effect in-between is a real physical effect. Another
essential result is that in the transition region the temperature is not a suitable external control 
parameter: The macrostate of the system cannot be adjusted by fixing the temperature. The better choice
is the system energy which is unfortunately difficult to control in experiments. Another direct 
consequence of the energetic ambiguity for a fixed temperature between $T_<$ and $T_>$ is that the
canonical interpretation is not suitable for detecting the backbending phenomenon. 

\begin{figure}
\includegraphics[height=.3\textheight]{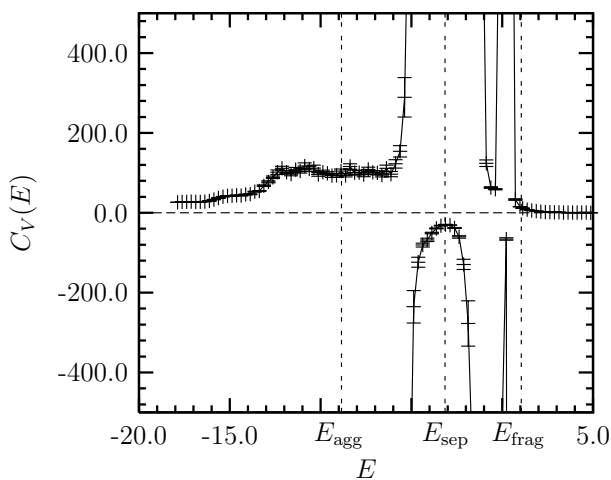}
\caption{\label{fig:agg:negC} 
Microcanonical specific heat $C_V(E)$ for the 2$\times$F1 complex. Note the negativity 
in the backbending regions~\cite{jbj1}.
}
\end{figure}

The most remarkable result is the negativity of the specific heat of the system in the
backbending region, as shown in Fig.~\ref{fig:agg:negC}.
A negative specific heat in the phase separation regime is due to the nonadditivity
of the energy of the two subsystems as the interaction between the chains 
is stronger than
the attractive inter-chain forces of the individual polymers. 
``Heating'' a \emph{large} aggregate would lead to the stretching of
monomer-monomer contact distances, i.e., the potential energy of an exemplified pair
of monomers increases, while kinetic energy and, therefore, temperature remain
widely constant. In a comparatively \emph{small} aggregate, additional energy leads
to cooperative rearrangements of monomers in the aggregate in order to reduce surface tension,
i.e, the formation of molten globular aggregates is suppressed.
In consequence, kinetic energy is transferred into potential energy and the temperature
decreases. In this regime, the aggregate becomes colder, although the total energy
increases ~\cite{jbj1}.

\begin{figure}
\includegraphics[height=.3\textheight]{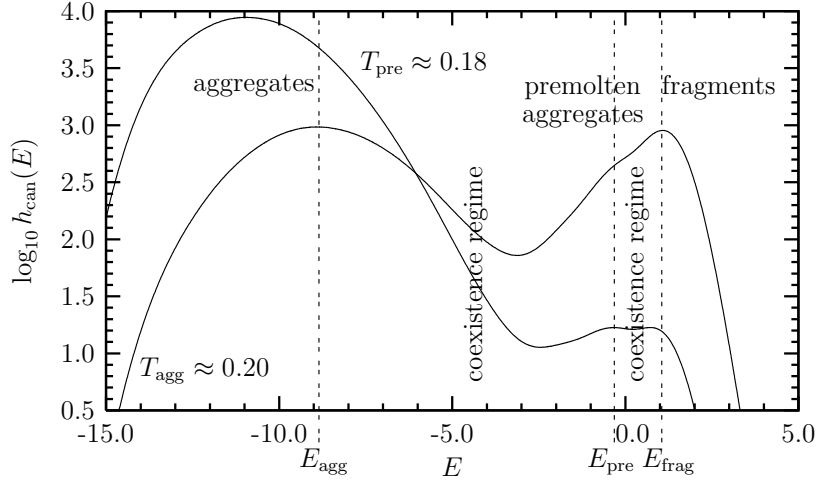}
\caption{\label{fig:agg:canhist} 
Logarithmic plots of the canonical energy histograms (not normalized) at $T\approx 0.18$ and 
$T\approx 0.20$, respectively.
}
\end{figure}

The precise microcanonical analysis reveals also a further detail of the aggregation transition. Close to
$E_\text{pre}\approx -0.32$, the $T^{-1}$ curve in Fig.~\ref{fig:agg:entropy} exhibits another
``backbending'' which signalizes a second, but unstable transition of the same type. 
The associated transition temperature $T_\text{pre}\approx 0.18$ is smaller than $T_\text{agg}$,
but this transition occurs in the energetic region where fragmented states dominate. Thus
this transition can be interpreted as the premelting of aggregates by forming intermediate states. 
These intermediate structures are rather weakly stable: The population of the premolten aggregates
never dominates. In particular, at $T_\text{pre}$, where premolten aggregates and 
fragments coexist, the population of compact aggregates is much larger. This can nicely be seen in
the canonical energy histograms at these temperatures plotted in Fig.~\ref{fig:agg:canhist}, where
the second backbending is only signalized by a small cusp in the coexistence region. Since both 
transitions are phase-separation processes, structure formation is accompanied by releasing latent
heat which can be defined as the energetic widths of the phase coexistence regimes, i.e., 
$\Delta Q_\text{agg}=E_\text{frag}-E_\text{agg}=T_\text{agg}[{\cal S}(E_\text{frag})-{\cal S}(E_\text{agg})]\approx 9.90$ 
and $\Delta Q_\text{pre}=E_\text{frag}-E_\text{pre}=T_\text{pre}[{\cal S}(E_\text{frag})-{\cal S}(E_\text{pre})]\approx 1.37$.
Obviously, the energy required to melt the premolten aggregate is much smaller than to dissolve a 
compact (solid) aggregate. 

For the comparison of the surface entropies, we use the definition~(\ref{eq:agg:ent}) of the entropy.
In the case of the aggregation transition, the surface entropy is 
$\Delta {\cal S}_\text{surf}^\text{agg} \approx \Delta S_\text{surf}^\text{agg}= H_S(E_\text{sep})-S(E_\text{sep})$,
where $H_S(E)\approx{\cal H}_{\cal S}(E)$ is the concave Gibbs hull of $S(E)$.
Since $H_S(E_\text{sep})=H_S(E_\text{frag})-(E_\text{frag}-E_\text{sep})/T_\text{agg}$ and 
$H_S(E_\text{frag})=S(E_\text{frag})$, the surface entropy is
\begin{equation}
\label{eq:agg:surfent}
\Delta S_\text{surf}^\text{agg}=S(E_\text{frag})-S(E_\text{sep}) - \frac{1}{T_\text{agg}}(E_\text{frag}-E_\text{sep}).
\end{equation}
Yet utilizing that the canonical 
distribution $h_\text{can}(E)=\int d\Gamma \,H_\text{can}(E,\Gamma;T)$
at $T_\text{agg}$ (shown 
in Fig.~\ref{fig:agg:canhist}) is $h_\text{can}(E)\sim g(E)\exp(-E/k_BT_\text{agg})$,
the surface entropy can be written in the simple and computationally convenient form~\cite{wj1}:
\begin{equation}
\label{eq:agg:surfent2}
\Delta S_\text{surf}^\text{agg}=k_B\ln\,\frac{h_\text{can}(E_\text{frag})}{h_\text{can}(E_\text{sep})}.
\end{equation}
A similar expression is valid for the coexistence of premolten and fragmented states at $T_\text{pre}$
The corresponding canonical distribution is also shown in Fig.~\ref{fig:agg:canhist}. Thus, we obtain
(in units of $k_B$) for the surface entropy of the aggregation transition 
$\Delta S_\text{surf}^\text{agg}\approx 2.48$ and for the premelting $\Delta S_\text{surf}^\text{pre}\approx 0.04$,
confirming the weakness of the interface between premolten aggregates and fragmented states.
\subsection{Aggregation transition in larger heteropolymer systems}

The statements in the previous section for the 
2$\times$F1 system are also, in general, valid for larger systems.
This is the result of computer simulations for systems consisting of three (in the
following referred to as 3$\times$F1) and four (4$\times$F1) 
identical peptides with sequence F1.
As it has already been discussed for the 2$\times$F1 system, there are also for the larger systems
no obvious signals for separate
aggregation and hydrophobic-core formation processes. Only weak activity in the energy
fluctuations in the temperature region below the aggregation transition temperature
indicates that local restructuring processes of little cooperativity 
(comparable with the discussion of the premolten aggregates in the discussion of the
2$\times$F1 system) are still happening. The strength of the aggregation transition is
also documented by the fact that the peak temperatures of energetic \emph{and} aggregation parameter
fluctuations are virtually identical for the multi-peptide systems, i.e., the aggregation 
temperature is $T_\text{agg}\approx 0.2$.

\begin{figure}
\includegraphics[height=.3\textheight]{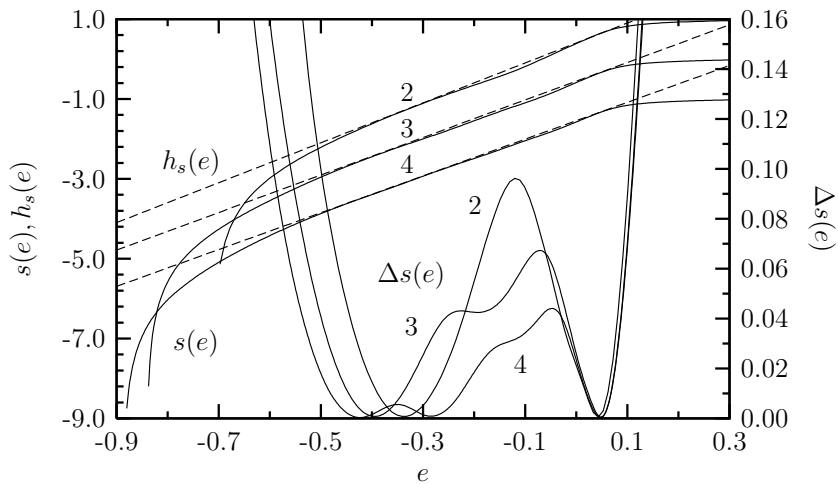}
\caption{\label{fig:agg:ent234} 
Microcanonical entropies per monomer $s(e)$, respective Gibbs constructions $h_s(e)$ (left-hand scale), and 
deviations $\Delta s(e)=h_s(e)-s(e)$ (right-hand scale) for 2$\times$F1 (labeled as 2), 3$\times$F1 (3), and 
4$\times$F1 (4) as functions of the energy per monomer $e$.
}
\end{figure}

In Fig.~\ref{fig:agg:ent234}, the microcanonical entropies per monomer $s(e)={\cal S}(e)/N_\text{tot}$ 
(shifted by an unimportant constant for clearer visibility) and the corresponding
Gibbs hulls $h_s(e) = {\cal H}_{\cal S}(e)/N_\text{tot}$ are shown for 2$\times$F1 (in the figure
denoted by ``2''), 3$\times$F1 (``3''), and 4$\times$F1 (``4''),
respectively, as  functions of the energy per monomer $e=E/N_\text{tot}$. 
Although the convex entropic ``intruder'' is apparent for larger
systems as well, its relative strength decreases with increasing number of chains. The slopes
of the respective Gibbs constructions determine the aggregation temperature~(\ref{eq:agg:tagg}) 
which are found to be $T_\text{agg}^{\text{3}\times\text{F1}}\approx 0.212$ and 
$T_\text{agg}^{\text{4}\times\text{F1}}\approx 0.217$. 

The existence of the interfacial boundary entails a transition barrier whose strength is characterized
by the surface entropy $\Delta {\cal S}_\text{surf}$. In Fig.~\ref{fig:agg:ent234},  the 
individual entropic deviations per monomer, $\Delta s(e)=\Delta {\cal S}(e)/N_\text{tot}$ 
are also shown and the maximum deviations, i.e., 
the surface entropies $\Delta {\cal S}_\text{surf}$ and relative surface entropies per monomer
$\Delta s_\text{surf}=\Delta {\cal S}_\text{surf}/N_\text{tot}$
are listed in Table~{\ref{tab:agg:lsys}}.
There is no apparent difference between the values of $\Delta {\cal S}_\text{surf}$ 
that would indicate a trend for a vanishing 
of the \emph{absolute} surface barrier in larger systems. However, the \emph{relative} surface entropy
$\Delta s_\text{surf}$
obviously decreases. Whether or not it vanishes in the thermodynamic limit cannot be decided
from our present results and is a study worth in its own right. 

It is also interesting that subleading effects increase 
and the double-well form found for 2$\times$F1 changes by higher-order effects,
and it seems that for larger systems the almost single-step aggregation of 2$\times$F1
is replaced by a multiple-step process.
\begin{table}
\caption{\label{tab:agg:lsys} Aggregation temperatures $T_\text{agg}$, surface entropies 
$\Delta {\cal S}_\text{surf}$, relative surface entropies per monomer $\Delta s_\text{surf}$,
relative aggregation and fragmentation energies per monomer, $e_\text{agg}$ and $e_\text{frag}$,
respectively, latent heat per monomer $\Delta q$, and phase-separation entropy per monomer
$\Delta q/T_\text{agg}$. All quantities for systems consisting of two, three, and four 
13-mers with AB sequence F1.}
\begin{tabular}{c|ccccccc}\hline
system & $T_\text{agg}$ & $\Delta {\cal S}_\text{surf}$ & $\Delta s_\text{surf}$ & 
$e_\text{agg}$ & $e_\text{frag}$ & $\Delta q$ & $\Delta q/T_\text{agg}$\\ \hline
2$\times$F1 & 0.198 & 2.48 & 0.10 & $-$0.34 & 0.04 & 0.38 & 1.92 \\
3$\times$F1 & 0.212 & 2.60 & 0.07 & $-$0.40 & 0.05 & 0.45 & 2.12 \\
4$\times$F1 & 0.217 & 2.30 & 0.04 & $-$0.43 & 0.05 & 0.48 & 2.21 \\ \hline
\end{tabular}
\end{table}

Not surprisingly,
the fragmented phase is hardly influenced by side effects and the rightmost minimum 
in Fig.~\ref{fig:agg:ent234} lies well at
$e_\text{frag}= E_\text{frag}/N_\text{tot}\approx 0.04-0.05$. Since the 
Gibbs construction 
covers the whole convex region of $s(e)$, the aggregation energy per monomer
$e_\text{agg}= E_\text{agg}/N_\text{tot}$ 
corresponds to the leftmost minimum and its value changes noticeably with the number of chains.
In consequence, the latent heat per monomer 
$\Delta q = \Delta Q/N_\text{tot}=T_\text{agg}[{\cal S}(E_\text{frag})-{\cal S}(E_\text{agg})]/N_\text{tot}$ that 
is required to fragment the aggregate increases from two to four chains in the 
system (see Table~\ref{tab:agg:lsys}). 
Although the systems under consideration are 
too small to extrapolate phase transition properties in the thermodynamic limit, 
it is obvious that the aggregation-fragmentation transition
exhibits strong similarities to condensation-evaporation transitions of colloidal systems. Given that,
the entropic transition barrier $\Delta q/T_\text{agg}$, which we see increasing with the number of
chains (cf.\ the values in Table~\ref{tab:agg:lsys}), would survive in the thermodynamic limit 
and the transition was 
first-order-like. More surprising would be, however, if the convex intruder would not 
disappear, i.e., if the absolute and relative surface entropies 
$\Delta {\cal S}_\text{surf}$ and $\Delta s_\text{surf}$ do not vanish. This is definitely
a question of fundamental interest as the common claim is that pure surface effects typically
exhibited only by ``small'' systems are irrelevant in the thermodynamic limit. This requires
studies of much larger systems. 

It should clearly be noted, however, that protein aggregates forming
themselves in biological systems often consist only of a few peptides and are definitely of small 
size and the surface effects are responsible for structure formation and are not unimportant
side effects. One should keep in mind that standard thermodynamics and the thermodynamic 
limit are somewhat theoretical constructs valid only for very large systems. 
The increasing interest in physical properties of small systems, in particular in conformational transitions
in molecular systems, requires in part a revision of dogmatic thermodynamic views. Indeed, by 
means of today's chemo-analytical and experimental equipment, effects like those described
in this chapter, should actually experimentally be verifiable as these are real physical
effects. For studies of the condensation of atoms, where a similar behavior occurs, 
such experiments have actually already been performed~\cite{schmidt1}.

\section{Summary}

The analyses in the previous sections have shown that it is indeed possible 
to reveal characteristic features of structure formation processes of
polymers, in particular proteins, by means of minimalistic coarse-grained
models. This is essential, as a generalized view of conformational transitions
occurring in folding and aggregation processes of molecular systems can 
only possess a solid basis, if a classification of generic features common to 
different systems enables the introduction of
suitable models on mesoscopic scales. 

Depending on the heteropolymer sequence,
typically two general transitions occur in heteropolymer
folding processes. One is the folding transition from random coils to compact globular conformations
common to all heteropolymers (i.e., little sequence-specific),
a finite-length analog to the collapse (or $\Theta$) transition known from homopolymers. 
The stability of the 
globular or intermediary (pseudo)phase of heteropolymers depends, however, 
strongly on the heteropolymer sequence. 
The second general transition at lower temperature (or worse solvent quality) 
is sort of a glassy transition as it results in the formation of the native conformation(s) with small
entropy. During this transition, the highly compact hydrophobic core is formed, surrounded by a
shell of polar residues which screens the core from the solvent. The 
kinetics of this transition strongly depends on the heteropolymer sequence. Hydrophobic-core
formation is typically a (``first-order-like'') phase separation process and
the sharpness and height of the free-energy barrier separating the hydrophobic-core and globular
(pseudo)phases are measures for the stability of the hydrophobic core. It is assumed that
a large set of the comparatively few functional bioproteins in nature exhibits such a large barrier 
preventing unfolding
into nonfunctional conformations. This is also one of the common arguments, why under 
physiological conditions
only a very small number among the possible protein sequences can be functional at all. 

Although the mesoscopic models are still extremely minimalistic,
we have found quite surprising characteristic folding features comparable to those of
real proteins.
Analyzing transition channels and free-energy landscapes based on a suitably defined 
similarity or ``order'' parameter, we identified folding behaviors which are known
from real proteins in a like manner: two-state folding with a single kinetic barrier 
and unique native state, folding towards the native fold 
through intermediates over different barriers, and metastability with different, almost degenerate,
native states. 

We have also discussed in detail thermodynamic properties of 
peptide aggregation processes. We compared small systems of different numbers of short peptides and 
investigated finite-size properties in the canonical and in the 
microcanonical ensemble. Each of these analyses has advantages.
Applying the canonical formalism reveals strong fluctuations in the vicinity
of the
aggregation transition which allow for a precise estimation of the aggregation transition
temperature for a finite system, but also for a finite-size scaling analysis
toward the system with infinitely many chains. 

But, analyzing the aggregation of a few peptides from the microcanonical
perspective uncovers an underlying physical effect, the backbending effect, which is 
largely ``averaged out'' in the canonical analysis. ``Backbending'' means that in the transition
region the caloric temperature decreases with increasing energy. This is due to surface
effects, additional energy does not lead to an increase of the caloric temperature; 
rather it is used to rearrange monomers in order to reduce surface tension at the expense of
entropy. In effect, the protein complex is getting colder. 
For an increasing number of peptides
in the system, we could show, however, that the effect becomes less relevant, although 
the latent heat increases and thus the first-order character of this phase-separation
process is getting stronger. Nonetheless, in biological aggregation processes typically 
only a few proteins are involved and thus the effect should be apparent. The ``physical reality'' 
of this effect has already been 
confirmed in atomic cluster formation experiments.
However, the experimental verification in polymeric systems is still pending.

One of the essential questions in aggregation processes among polymers
is, how the mutual influence induces 
conformational changes. Two potential scenarios leading to the formation of 
complexes are conceivable. If the external force is attractive, but weaker than the intrinsic,
intermonomeric forces that form the polymer or protein conformation, the proximity of
an attractive polymer or substrate is not sufficient to
refold the polymer and the aggregation is a simple docking process. Unless
the match is perfect, the binding force that holds the compound together is rather weak.
On the other hand, if the external force entails refolding of the polymer, it can better adapt
to the target structure (e.g., a crystalline substrate), or, if both partners experience 
conformational changes, a new, highly compact compound can form. 
In this so-called coupled folding-binding
process, the binding force is typically stronger than in the docking case. In our aggregation
study of a few short peptides, we observed such a behavior. 

Our results were mainly obtained
by means of sophisticated generalized-ensemble chain-growth and Markov chain Monte Carlo
computer simulations, partly newly developed or generalized for these purposes. It is a
non-negligible fact that even with today's equipment computer simulations of 
polymers, in particular, proteins, are extremely demanding and efficient algorithms 
are required. Despite the enormous progress in protein research in the past few years,
it will remain one of the biggest scientific future challenges to uncover the principal
secrets of cooperative conformational activity in structure formation processes of proteins.

\begin{theacknowledgments}
We thank S.\ Schnabel and C.\ Junghans for the collaboration in the mesoscopic protein folding and
aggregation projects. 
This work is partially supported by the DFG (German Science Foundation) under Grant
No.\ JA 483/24-1/2, the Leipzig Graduate School of Excellence 
``BuildMoNa~-- Building with Molecules and Nano-objects'',
and the German-French DFH-UFA PhD College ``Statistical Physics of Complex Systems''
under Grant No.\ CDFA-02-07.
Supercomputer time at the John von Neumann Institute for Computing (NIC),
Forschungszentrum J\"ulich, is acknowledged (Grant No.\ hlz11).
\end{theacknowledgments}

\end{document}